# A FRAP model to investigate reaction-diffusion of proteins within a bounded domain: a theoretical approach

George D.Tsibidis[*]

Institute of Electronic Structure and Laser, Foundation for Research and Technology, P.O.Box 1385, Vassilika Vouton, 71110 Heraklion, Crete, Greece

**Abstract**

Temporally and spatially resolved measurements of protein transport inside cells provide important clues to the functional architecture and dynamics of biological systems. Fluorescence Recovery After Photobleaching (FRAP) technique has been used over the past three decades to measure the mobility of macromolecules and protein transport and interaction with immobile structures inside the cell nucleus. A theoretical model is presented that aims to describe protein transport inside the nucleus, a process which is influenced by the presence of a boundary (i.e. membrane). A set of reaction-diffusion equations is employed to model both the diffusion of proteins and their interaction with immobile binding sites. The proposed model has been designed to be applied to biological samples with a Confocal Laser Scanning Microscope (CLSM) equipped with the feature to bleach regions characterised by a scanning beam that has a radially Gaussian distributed profile. The proposed model leads to FRAP curves that depend on the on- and off-rates. Semi-analytical expressions are used to define the boundaries of on- (off-) rate parameter space in simplified cases when molecules move within a bounded domain. The theoretical model can be used in conjunction to experimental data acquired by CLSM to investigate the biophysical properties of proteins in living cells.

[*] Corresponding author, Tel: 00302810391912, Fax: 0030-2810391569

E-mail address: tsibidis@iesl.forth.gr

# 1 Introduction

Structural and regulation requirements of nuclear processes such as DNA replication, transcription and functional architecture and dynamics are as yet not well understood. Although, important advances towards a better understanding have been accomplished, deeper questions in regard to the interaction of nuclear proteins and dynamics still remain unanswered. Temporally and spatially resolved measurements of protein transport inside cells provide important clues to the functional architecture and dynamics of biological systems. They can provide information on how proteins interact and how they function which, in turn, allows to develop a more accurate picture of the underlying environment within the cell and measure conformational changes (for a review, see (Lippincott-Schwartz and others 2001; Phair and Misteli 2001).

Fluorescence Recovery After Photobleaching (FRAP) has become a very useful tool to measure the translational components of sub-cellular systems (for a review, see (Carmo-Fonseca and others 2002; Carrero and others 2003; Sprague and McNally 2005). The technique has been widely applied to systems expressing GFP fusion proteins for the study of a variety of problems including measuring diffusion coefficients of macromolecules in solution (Axelrod and others 1976; Soumpasis 1983), probe diffusion in tissues (Berk and others 1993), diffusion in cytoplasm (Luby-Phelps and others 1987), transport in the endoplasmic reticulum and Golgi (Lippincott-Schwartz and others 2001) and transport inside the cell nucleus (Patterson and Lippincott-Schwartz 2002).

The availability of confocal scanning laser microscopes (CLSM) opened new possibilities for photobleaching methods (Scholz and others 1988) as they are often equipped with the feature to bleach selectively regions in the sample with high spatial resolution. Probing of specific areas of the cell allows an enhanced understanding of biological sub-cellular processes. In principle, the majority of functional proteins inside living cells are not free to diffuse but they rather bind and unbind transiently to other species forming multi protein complexes (Alberts 1998; Dundr and others 2002). A few FRAP models exist which in conjunction with a CLSM can be used to investigate systematically diffusion and binding behaviour of nuclear proteins. All models assumed that a set of reaction-diffusion equations has to be used to determine the contributions of protein interaction and diffusion on fluorescence redistribution. Most of the models assumed one dimensional (Carrero and others 2004a; Carrero and others 2004b; Carrero and others 2003; Hinow and others 2006; Tardy and others 1995), a two dimensional (Braga and others 2007; Sprague and others 2004) or a three-dimensional (Beaudouin and others 2006; Sprague and others 2006) process in bounded domains. Most of the models assumed a circular uniform bleaching profile for the laser beam of the CLSM.

In general, the bleaching profile of CLSM has a Gaussian profile which in a number of previous studies has been approximated with a uniform circular profile. As stated in a previous work (Tsibidis and Ripoll), this simplification is valid only if the radial resolution of the laser bleaching beam is much smaller than the size of the bleached spot (Braeckmans and others 2003). By contrast, for comparable sizes, this assumption yields erroneous results that lead an overestimation of the rate constants.

We aim to analyse in a systematic way the effect of a bleaching profile of a Gaussian form (i.e. non uniform bleaching). In principle, the form of the bleaching profile should not affect interaction of biomolecules with binding sites inside the nucleus. Contributions to the form of the recovery curve due to the different type should nevertheless be taken into consideration. For the sake of simplicity, a low numerical aperture (NA) of the objective of CLSM will be considered since it causes the bleaching of an approximately cylindrical region. As a result, diffusion of proteins can be regarded to be two-dimensional on the focal plane (Tsibidis and Ripoll).

In this study, we will develop a model that aims to describe reaction-diffusion processes inside a cell nucleus that is taken to be a bounded domain (of radius *b*) which restricts diffusion of nuclear proteins out of the boundary. We apply our model to investigate the diffusion of GFP-tagged proteins with diffusion coefficient $Df$=30μm$^2$/sec after a spot size of radius *w*=1μm was bleached using a CLSM with a low NA. Expressions are derived for the fluorescence recovery curve that contain parameters such as the bleach depth and the half width of the intensity of the bleaching beam which can be calculated by analysing experimental data. In order to examine the binding of the proteins to immobile sites, various regimes of pure and constrained diffusion were investigated by changing the values of the associating and dissociating rates. The regimes were distinguished by different time scales. We believe that our analysis can offer an efficient tool for fitting experimental data that will lead to obtain significant insight into the quantitative characterisation of nuclear proteins undergoing binding-unbinding events.

## 2 Model formulation

*2.1 Bleaching of a disk with CLSM*

It is assumed that bleaching is described by an irreversible first-order reaction and we take the bleaching PSF, $I_{bleach}(r)$ to be radially Gaussian distributed according to the formula

$$I_{bleach}(r) \propto \exp(-r^2/r_0^2) \qquad (1)$$

where $r_o$ is the radial resolution of the beam. If photobleaching is performed on a nucleus (Figure 1A) using a high numerical aperture (NA), fluorescence depletion is the same on average in every focal plane (Tsibidis and Ripoll) and the concentration of the fluorescent proteins as a function of radial distance is (Axelrod and others 1976; Blonk and others 1993; Braga and others 2004)

$$C_{unbleached}(r) = C_i \exp(-K \exp(-r^2/r_0^2)) = C_i \sum_{n=0}^{\infty} \frac{(-K)^n}{n!} \exp(-2nr^2/r_0^2) \qquad (2)$$

where *K* denotes the bleach constant. $C_i$ represents the uniform prebleach fluorescing species concentration. In Figure 1B, the nucleus of a cell is approximately illustrated as a cylinder of radius *b* and height *H*.

*2.2 A reaction-diffusion model for FRAP*

We assume that a reaction of the form

$$F + B \leftrightarrow C \qquad (3)$$

describes the binding of free diffusing proteins with binding sites where *F* represents free proteins, *B* represents vacant and immobile binding sites and *C* represent bound complexes. Free proteins bind and unbind to binding sites with rate constants $k_b$ and $k_u$, respectively. The average time for diffusion between binding events is $t_d=1/k_b$, while the average residency time of proteins in bound form is $t_r=1/k_u$. Binding mechanisms can be visualised clearly by fusing *F* with fluorescently tagged proteins. Unlike the *C* species which becomes fluorescent when binding reaction occurs, *B* are always non-fluorescent. The bound species is assumed not to diffuse although this assumption may not be always true for some complexes (Sprague and others 2006). The unbound proteins are expected to diffuse freely and photobleaching is performed on a fluorescent population that has reached a uniform-steady state distribution. Additionally, the distribution of the binding sites is considered to be homogeneous and remain constant during the fluorescence recovery.

Based on the above assumptions, the model that describes binding and unbinding of bound and free proteins can be written mathematically as a system of reaction-diffusion equations:

$$\frac{\partial f}{\partial t} = D_f \nabla_r^2 f - K_b f + k_u c$$
$$\frac{\partial c}{\partial t} = K_b f - k_u c \qquad (4)$$

where *f* and *c* represent the concentration of fluorescent *F* and *C*, respectively, $D_f$ is the diffusion coefficient of free proteins *F* and $K_b \equiv k_b B_o$. The subscript *r* in the Laplacian operator indicates that all axial terms have been removed from the equation due to the two dimensional character of the process. Given the Gaussian form of the bleaching profile, the initial conditions are:

$$f(r,t=0) = \frac{1}{1+\gamma} C_i \exp(-K \exp(-2r^2/r_o^2))$$
$$c(r,t=0) = \frac{\gamma}{1+\gamma} C_i \exp(-K \exp(-2r^2/r_o^2)) \qquad (5)$$

where $\gamma = K_b/k_u$, and $F_o \equiv 1/(1+\gamma)$ and $C_o \equiv \gamma/(1+\gamma)$ represent the proportions of fluorescent free and bound populations, respectively.

In most cases, the presence of membranes in cells influences the diffusion of proteins and thereby the infinite domain expressions presented in other studies (Sprague and others 2004; Tsibidis and Ripoll) do not suffice to provide an accurate estimate of the fluorescence recovery. For the sake of simplicity, we assume firstly that the bleached area is located around the centre of the nucleus. Although, the same set of reaction-diffusion equations govern the dynamics of the system, Neumann boundary conditions should be introduced as constraints corresponding to the fact that the flux of fluorescent molecules outside the nuclear membrane should be zero

$$\left.\frac{\partial f}{\partial r}\right|_{r=b} = \left.\frac{\partial c}{\partial r}\right|_{r=b} = 0 \tag{6}$$

while the initial conditions are given by Eq.5. In Section B in Supplementary Material, the reaction-diffusion equations are solved taking into consideration the initial and boundary conditions and the fluorescence intensity in the *p*-space is expressed by the following equation

$$F(p) = \frac{1}{p} + \frac{2F_o}{w^2 D_f}\left(1 + \frac{K_b}{p+k_u}\right)^2 \sum_{n=1}^{\infty}\frac{(-K)^n}{n!}B(q,A,w,n,r_o) +$$

$$\frac{C_o}{p+k_u}\frac{1}{2}\left(\frac{r_o}{w}\right)^2 \sum_{n=1}^{\infty}\frac{(-K)^n}{n!n}\left(1-\exp(-2n(w/r_o)^2)\right),$$

$$B(q,A,w,n,r_o) = \left[\int_0^w drr\left(\int drr\left[K_0(qr)+I_0(qr)\frac{K_1(qb)}{I_1(qb)}\right]\exp(-2n(r/r_o)^2)\right)\right|_{r=A} I_0(qr) - \tag{7}$$

$$\int_0^w drrI_0(qr)\int drrK_0(qr)\exp(-2n(r/r_o)^2) + \int_0^w drrK_0(qr)\int drrI_0(qr)\exp(-2n(r/r_o)^2)\right],$$

$$q^2 = \frac{p}{D_f}\left(1+\frac{K_b}{p+k_u}\right)$$

It is obvious that Eq.7 reduces to the expression derived for diffusion in an infinite domain (Tsibidis and Ripoll) for large values of *b* (as $b\rightarrow\infty$, $K_1(qb)/I_1(qb) \rightarrow 0$) where *p* is the Laplace variable and $I_o$, $I_1$ ($K_o$, $K_1$) are the modified Bessel functions of first (second) kind and *A* is a cut-off value (see Section B in Supplementary Material). Due to its complex form, to the best of our knowledge, an analytical expression of *F(p)* cannot be obtained and a numerical method has to be pursued to evaluate the integrals. Time evolution of the FRAP recovery curve is computed by calculating the inverse Laplace transform of *F(p)* by means of the Matlab algorithm *invlap.m* to obtain the FRAP recovery curve. The parameters $t_r$ and $t_d$ operate as the timescales that can determine protein dynamics reduction to simpler scenarios (Tsibidis and Ripoll):

*I. Pure and effective diffusion scenarios*

Unlike the pure and effective diffusion simplified scenarios in the case of protein movement in an infinite domain (Axelrod and others 1976; Braeckmans and others 2003; Braga and others 2004; Carmo-Fonseca and others 2002; Soumpasis 1983; Sprague and others 2004; Tsibidis and Ripoll), an analytic expression in a closed form when the domain is bounded cannot be obtained. The expression derived, though, for the full model can simply be reduced to simpler expressions by noting that $C_o=0$ and that $K_b/(p+k_u)\approx 0$ (pure diffusion) and $p+k_u\approx k_u$. Pure diffusion expression in bounded domains can be derived by following a variable separation procedure (Crank 1975) for the spatial and temporal parts of the pure diffusion equation. For diffusion of fluorescent molecules in a cylindrical volume bounded at $r=b$, we obtain the following expression for the concentration of fluorescent material (Crank 1975)

$$f(r,t) = C_i - \frac{2}{b^2}\int_0^b F(r)rdr - \frac{2}{b^2}\sum_{n=1}^{\infty}\exp(-D_f t\alpha_n^2)\frac{J_0(\alpha_n r)}{(J_0(\alpha_n b))^2}\int_0^b rF(r)J_0(\alpha_n r)dr$$

where $J_o$ are the zero-th order Bessel functions of the first kind and $\alpha_n$ are values for which the first derivative of the first order Bessel functions of the first kind, $J_1(\alpha_n b)$ equals zero (this results from the boundary condition). By recalling that the initial concentration of the bleached biomolecules at $t=0$ $F(r)=C_{bleached}$, the above expression turns into (noting that $J_1(\alpha_n b)=0$)

$$f(r,t) = \frac{r_o^2 C_i}{2b^2}\sum_{m=1}^{\infty}\frac{(-K)^m}{mm!}\left(1-\exp(-2mb^2/r_o^2)\right) +$$

$$\frac{2C_i}{b^2}\sum_{n=1}^{\infty}\exp(-D_f t\alpha_n^2)\frac{J_0(\alpha_n r)}{(J_0(\alpha_n b))^2}\int_0^b rJ_0(\alpha_n r)\left[\sum_{m=1}^{\infty}\frac{(-K)^m}{m!}\exp(-2mr^2/r_o^2)\right]dr \quad (8)$$

In Supplementary Material, we have also derived an expression in an integral form that describes the spatio-temporal distribution of fluorescent biomolecules for pure diffusion (Eq.SM.7) and the normalised fluorescence intensity (Eq.SM.14). The spatial distribution of fluorescent biomolecules was computed to test our model in the idealised case of pure diffusion: (i) by applying Eq.8 and (ii) by calculating the inverse Laplace transform of Eq.SM.7

$$-\frac{F_o}{D_f}\left(1+\frac{K_b}{p+k_u}\right)\sum_{n=1}^{\infty}\frac{(-K)^n}{n!}B(q,A,w,n,r_o),$$

$$B(q,A,w,n,r_o) = \left[\int drr\left[K_0(qr) + I_0(qr)\frac{K_1(qb)}{I_1(qb)}\right]\exp(-2n(r/r_o)^2)\Big|_{r=A} I_0(qr) - \right. \quad (9)$$

$$\left. I_0(qr)\int drrK_0(qr)\exp(-2n(r/r_o)^2) + K_0(qr)\int drrI_0(qr)\exp(-2n(r/r_o)^2)\right],$$

Figure 1C demonstrates the agreement of the results yielded by the two methods for the following parameter values: $D_f=1\mu m^2/sec$, $K_b=10sec^{-1}$, $k_u=10^4 sec^{-1}$. The excellent agreement of our model with well-established results for a pure diffusive behaviour constitutes a good, initial test for our model. Similar expressions are valid for effective diffusion which occurs when reaction is very fast compared to the time required for proteins to diffuse. In that case, the expression that describes the protein dynamics is the same as for pure diffusion with the substitution $D_f \rightarrow D_{eff}$.

*II. Reaction dominant scenario*

When diffusion occurs so fast that it is essentially undetectable during the FRAP experiment, the following analytical expression describes the regime where binding reaction dominates (Eq.A.3):

$$F(t) = 1 + \frac{1}{2}\left(\frac{r_o}{b}\right)^2 \sum_{n=1}^{\infty} \frac{(-K)^n}{n!n}\left(1 - \exp(-2n(b/r_o)^2)\right) +$$

$$\frac{\gamma}{1+\gamma}\left[\frac{1}{2}\left(\frac{r_o}{w}\right)^2 \sum_{n=1}^{\infty} \frac{(-K)^n}{n!n}\left(1 - \exp(-2n(w/r_o)^2)\right) - \right.$$

$$\left. \frac{1}{2}\left(\frac{r_o}{b}\right)^2 \sum_{n=1}^{\infty} \frac{(-K)^n}{n!n}\left(1 - \exp(-2n(b/r_o)^2)\right)\right]\exp(-k_u t) \qquad (10)$$

We note the dependence of the fluorescence intensity to the size of the bleached area and that for large values of the boundary ($b \rightarrow \infty$), the expression reduces to the form of the solution for an infinite domain (Tsibidis and Ripoll).

## 3  Results and Discussion

A systematic analysis was performed to investigate intracellular protein transport in the presence of binding sites to investigate protein dynamics in living cells. Compared to previous analytical models that aimed to identify and quantify basic interactions that regulate sub-cellular processes, we presented an 'improved' physical model to simulate and describe complex and dynamic biological processes. The approach offers distinct advantages considering that for bleached spot sizes comparable to the resolution of the beam, a Gaussian profile should describe better fluorescence depletion.

In the following, the full model is examined against the three idealised cases (pure diffusion, effective diffusion and reaction dominant) to investigate the rate constant space ($K_b, k_u$) in which protein behaviour can be simplified. To investigate the various behaviours, we considered GFP-tagged proteins and we set the diffusion coefficient to $D_f = 30 \mu m^2/sec$, the radial resolution to $r_o = 1.2 \mu m$, the bleached spot radius to $w = 1 \mu m$ and the bleach constant to $K=4$. This choice of parameters has been used previously in experiments to compute nucleoplasmic diffusion coefficients for SF2/ASF, fibrillarin and HMG-17 (Phair and Misteli 2000).

To investigate the intracellular diffusion and binding interactions, we set the diameter of nucleus to $2b = 12 \mu m$. Unlike diffusion in an unbounded domain, the intensity for $F$ and $C$ populations will never recover to their pre-bleach values $F_o$ and $C_o$, respectively (see Figure 2). A solution is derived by the application of the Laplace transformation (Eq.7).

To examine thoroughly and validate the proposed FRAP model and the influence of the rate constant parameters, a study of the FRAP recovery curves was conducted in the parameter space of both the binding and unbinding rates in the interval [$10^{-6}$sec$^{-1}$, $10^{6}$sec$^{-1}$]. For the sake of simplicity, a logarithmic space was used and a grid of all possible values for $K_b$ and $k_u$ was built in $10^{0.3}$ increments. For every pair of values, the FRAP recovery curve was computed considering the Laplace inverse of Eq.7 and each of the three simplified scenarios were used to generate the curves for each corresponding case. All calculations were performed using Matlab and for every pair of values, computations and comparisons with analytical solutions required about 20 secs. We used $n_{max}$=20 as the number of terms considered when an infinite series had to be calculated. There are values for which the full model can be approximated quite satisfactorily with one of the simplified cases: pure diffusion (Figure 2A), effective diffusion (Figure 2B) or reaction dominant case (Figure 2C). This suggests that full reaction-diffusion model reduces to simpler cases. Additionally, there exist values for which none of the idealised scenarios appear to be adequate to fit the full model (Figure 2D). The goodness of agreement between the two solutions was provided by the sum of the square root of residuals of the two recovery curves spot.

To determine the regions where the limiting cases hold true we chose a threshold value $S$=0.1. Figure 3A displays the regions where the simplified and the full model describe protein transport for particular combinations of the rate constants. Our method failed to compute FRAP recovery curve in the upper part region of Figure 3A and thereby a numerical approach was used by means of the *pdepe* Matlab function. The employment of this method indicates that the region in the upper left part of the parameter space is essentially part of the effective diffusion regime. A similar approach was also implemented for an even smaller bounded domain ($b$=4μm) to test whether the pattern changed substantially. It turns out that the pattern of the rate constants is similar to that illustrated in Figure 3A, suggesting that the regimes always occupy the same regions regardless of the size of the boundary.

The comparison of the fluorescence curves for infinite and bounded domains emphasises the role of the distance of the nuclear membrane from the bleached area in the form of the recovery intensities. For three barrier sizes (i.e. $b$=6, 8 and 30μm), fluorescence intensity curves in the bleaching spot were produced and were tested against the theoretical results in the absence of a barrier. In Figure 3B, the dependence of the theoretical recovery on the size of the bounded domain is demonstrated for a system that behaves according to the full-reaction model ($K_b$=100sec$^{-1}$ and $k_u$=10$^{-2.5}$sec$^{-1}$). Similar curves can be produced exhibiting the same pattern for the idealised cases. All of the curves correspond to the same pair of reaction rates and the further the boundary is located at, the better fitting with the infinite domain curve is, as expected. The reduction on the final value of fluorescence is due to the insufficient amount of fluorescence inside the domain to outweigh the loss of fluorescence due to bleaching. For large values of the barrier size ($b$=30μm), the results are similar to the theoretical results in the infinite domain case. It appears that during the first phase of the recovery, the size of the membrane does not play a significant role. As a result, expressions which are valid for the infinite domain can be perfectly used to fit experimental data that describe fluorescence recovery for bounded domains (at early timepoints) producing the same quantitative information about the interaction with immobile structures. This type of behaviour is consistent with a previous analysis of the effects of an impermeable boundary on pure diffusion of proteins (Carrero and others 2003). We have generalised the findings for 2D movement of proteins taking into account the special form of the beaching profile.

In the above analysis, we assumed that the spot is in the centre of the nucleus. To estimate the fluorescence recovery when bleached spot is displaced from the nucleus centre, the system of reaction-diffusion equations Eq.4 is used but with the replacement of the Laplacian operator $\nabla_r^2 \to \nabla_{x,y}^2$. Without loss of generality and for the sake of a simpler formulation, we have assumed that the bleached spot has been displaced along only the *x*-direction. We have chosen a bleached spot displaced 4μm from the nucleus centre. To the best of our knowledge, an analytical expression cannot be obtained and a numerical calculation of the reaction-diffusion equations was performed with a finite element method that is employed by the use of the powerful and flexible *pdetool* of Matlab: A refined mesh with 2145 nodes and 4160 triangles inside the nucleus yielded very good fitting (Figure 3C) for no displacement of the bleached spot (*d*=0.0μm) which constitutes a test for the choice of the mesh size and FEM method. The values used for the rate constants were $K_b=10^2 sec^{-1}$, $k_u=10^{-2.5} sec^{-1}$. By contrast, it turns out that when the bleached spot resides closer to the boundary (*d*=1.5 and 3μm), diffusion of the proteins is affected by the proximity to the boundary leading to a slower recovery except from the earlier timepoints (Figure 3C). Previous analysis (Carrero and others 2003; Sprague and others 2006) yielded similar results for a uniform circular bleaching profile.

This type of behaviour holds true for all values of the binding parameters with the exception of those that belong to the reaction dominant regime. This difference indicates that the boundary appears to influence the diffusive character of the process. Since diffusion occurs very fast before the system reaches an equilibrium, the effect of the boundary is very small for processes described by a reaction dominant scenario.

It is important to note that the Gaussian bleaching profile induces a 'tail' along a small distance from the bleached spot. As a result, a special attention is required regarding the proximity of the spot to the membrane. If it is taken to be very close to the boundary, the initial bleaching profile will not be symmetric around the centre of the bleached spot. The employment of our approach would yield erroneous results since it would assume a symmetrical bleaching without an azimuthal dependence.

One important consequence of our model which is not present for a uniform circular bleaching profile, is the dependence of fluorescence recovery on the spot size when the diffusion is very fast and reaction dominates. If the bleached spot size changes and the radial resolution changes in a manner that $r_o/w$ is constant, fluorescence recovery curve remains the same. These results are consistent with experimental data after depletion of ATP via sodium azide treatment (Tang and DeFranco 1996).

A critical assumption of our approach was that the binding sites are uniformly distributed in the sample. Although this appears to be an oversimplification that might lead to wrong results, there are cases where this assertion is valid. More specifically, in experiments where the transport of GFP-histone H1 in the nucleus was investigated (Th'ng and others 2005), the binding sites of chromatin are distributed uniformly throughout the nucleus. Additionally, the distribution of histone shows there is a uniform accumulation of the protein inside nuclei without the formation of clusters.

## 5 Conclusions

In this work, we focused on the study of a particular type of biological processes: the protein transport inside mammalian cells provided they are confined to move within bounded domains. CLSM was used to photobleach a circular region assuming that the bleaching profile of the beam has a radially Gaussian distribution and then, interaction of proteins with binding sites was analysed using a mathematical model. We believe that our present analysis provides the theoretical basis for the determination of the factors related to protein transport and interaction. It offers an efficient tool for fitting experimental data that will lead to gain significant insight into the quantitative characterisation of nuclear proteins undergoing binding-unbinding events. It is hoped that this work will help to enrich our understanding of protein function that underlies complex dynamics of cytoskeleton and nuclear matrix.

**Acknowledgements**

G.D. Tsibidis acknowledges support from the EU Coordination Action Project ENOC 022496.

**Appendix**

*Reaction-dominant reduction of reaction-diffusion equation*

When the system lies in the reaction-dominant regime, diffusion of molecules is a very fast process. As a result, it is the concentration of fluorescent bound molecules that has a time-dependence. By contrast, the concentration of free molecules reaches fast an equilibrium. The total number of free fluorescent molecules inside the circle of radius $b$ is $\pi b^2 f$, where $f$ is the concentration of free fluorescent molecules inside the nucleus. This number equals to the number of molecules that are not bleached

$$\int_0^b dr\, 2\pi r F_o C_i \exp(-K \exp(-2r^2/r_o^2)) = \pi b^2 f \Rightarrow$$

(A.1)

$$f = F_o C_i \left[1 + \sum_{n=1}^{\infty} \frac{(-K)^n}{2n!\,n} \left(\frac{r_o}{b}\right)^2 \left(1 - \exp(-2nb^2/r_o^2)\right)\right]$$

If we substitute this expression into the second equation of Eq.4, we obtain the concentration of the fluorescent bound molecules

$$c(t) = C_o C_i + \frac{C_o C_i}{2}\left(\frac{r_o}{b}\right)^2 \sum_{n=1}^{\infty} \frac{(-K)^n}{n!\,n}\left(1 - \exp(-2n(b/r_o)^2)\right) +$$

$$C_o C_i \left[\frac{1}{2}\left(\frac{r_o}{w}\right)^2 \sum_{n=1}^{\infty} \frac{(-K)^n}{n!\,n}\left(1 - \exp(-2n(w/r_o)^2)\right) - \right.$$

(A.2)

$$\left. \frac{1}{2}\left(\frac{r_o}{b}\right)^2 \sum_{n=1}^{\infty} \frac{(-K)^n}{n!\,n}\left(1 - \exp(-2n(b/r_o)^2)\right)\right] \exp(-k_u t)$$

and if we recall that $F_o + C_o = 1$ and $C_o = \gamma/(1+\gamma)$, we finally get

$$F(t) = 1 + \frac{1}{2}\left(\frac{r_o}{b}\right)^2 \sum_{n=1}^{\infty} \frac{(-K)^n}{n!\,n}\left(1 - \exp(-2n(b/r_o)^2)\right) +$$

$$\frac{\gamma}{1+\gamma}\left[\frac{1}{2}\left(\frac{r_o}{w}\right)^2 \sum_{n=1}^{\infty} \frac{(-K)^n}{n!\,n}\left(1 - \exp(-2n(w/r_o)^2)\right) - \right.$$

(A.3)

$$\left. \frac{1}{2}\left(\frac{r_o}{b}\right)^2 \sum_{n=1}^{\infty} \frac{(-K)^n}{n!\,n}\left(1 - \exp(-2n(b/r_o)^2)\right)\right] \exp(-k_u t)$$

**List of Figures**

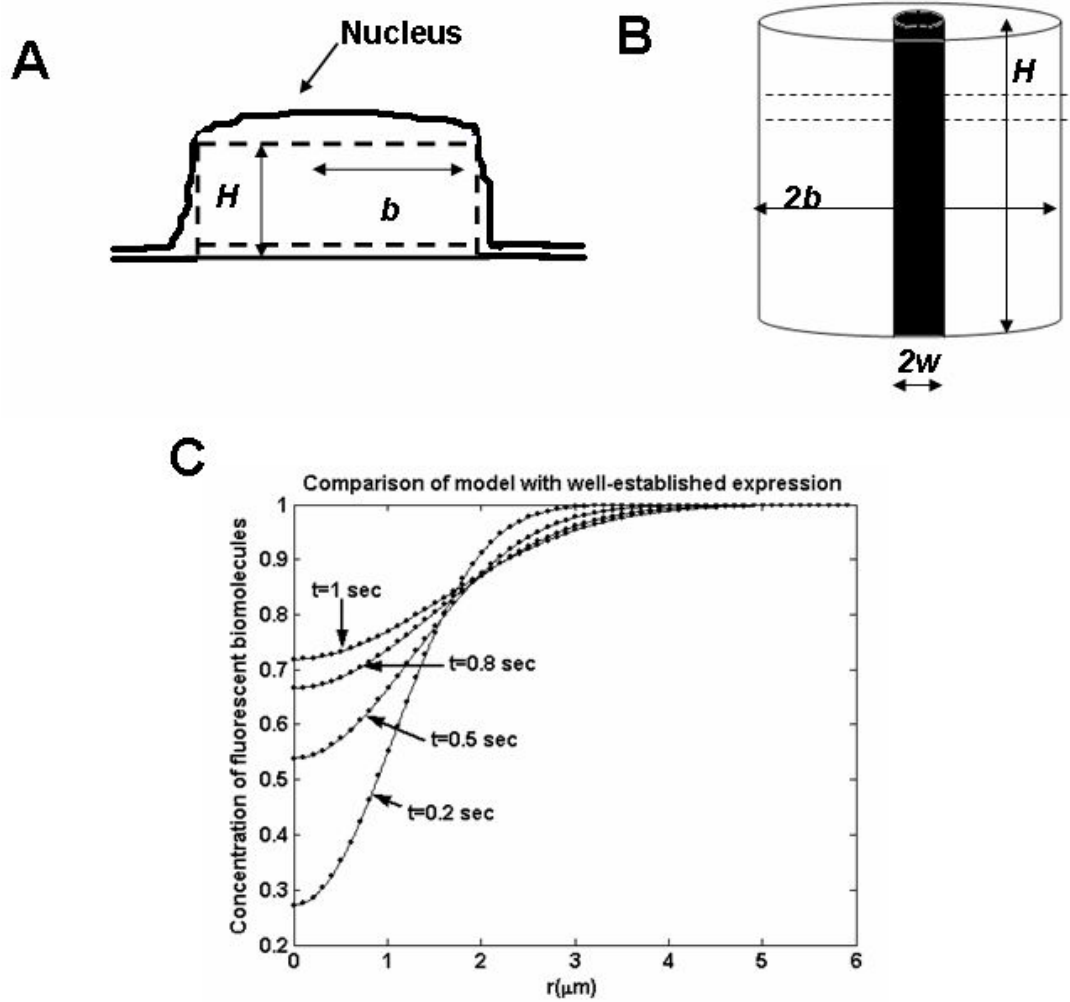

Figure 1: Geometry of the biological system. (A) Cell nucleus of height $H$ and radius $b$. (B) A cylinder of height $H$ and diameter $2b$, approximating the cell nucleus and a circular bleached region of radius $w$. (C) Comparison of results derived from Eq.8 (*curves*) and Eq.9 (*dots*) at timepoints t=0.2,0.5,0.8 and 0.1secs following values: $D_f$ =1µm$^2$/sec, $K_b$=10sec$^{-1}$, $k_u$=10$^4$sec$^{-1}$.

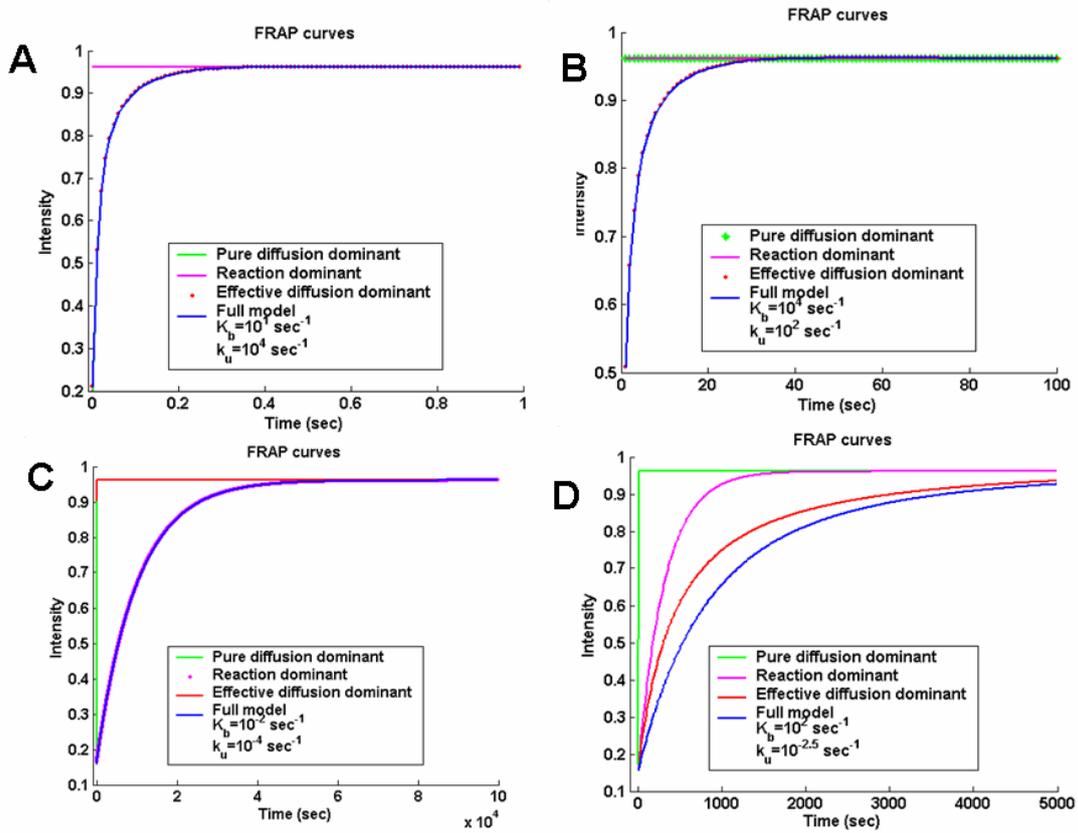

Figure 2: FRAP curves for various values of the binding and unbinding rates $K_b$, $k_u$ for diffusion in a bounded domain. Full model and curves that are valid in simplified regimes are plotted to test whether the pair of values mark a specific behaviour. (A) If $K_b=10^1 \text{sec}^{-1}$, $k_u=10^4 \text{sec}^{-1}$, the system can be described satisfactorily by a pure diffusion model. (B) If $K_b=10^4 \text{sec}^{-1}$, $k_u=10^2 \text{sec}^{-1}$, the system can be described satisfactorily by an effective diffusion model. (C) If $K_b=10^{-2} \text{sec}^{-1}$, $k_u=10^{-4} \text{sec}^{-1}$, the system can be described satisfactorily by a reaction-dominant model. (D) If $K_b=10^2 \text{sec}^{-1}$, $k_u=10^{-2.5} \text{sec}^{-1}$, no simplified model can approximate the behaviour of the system.

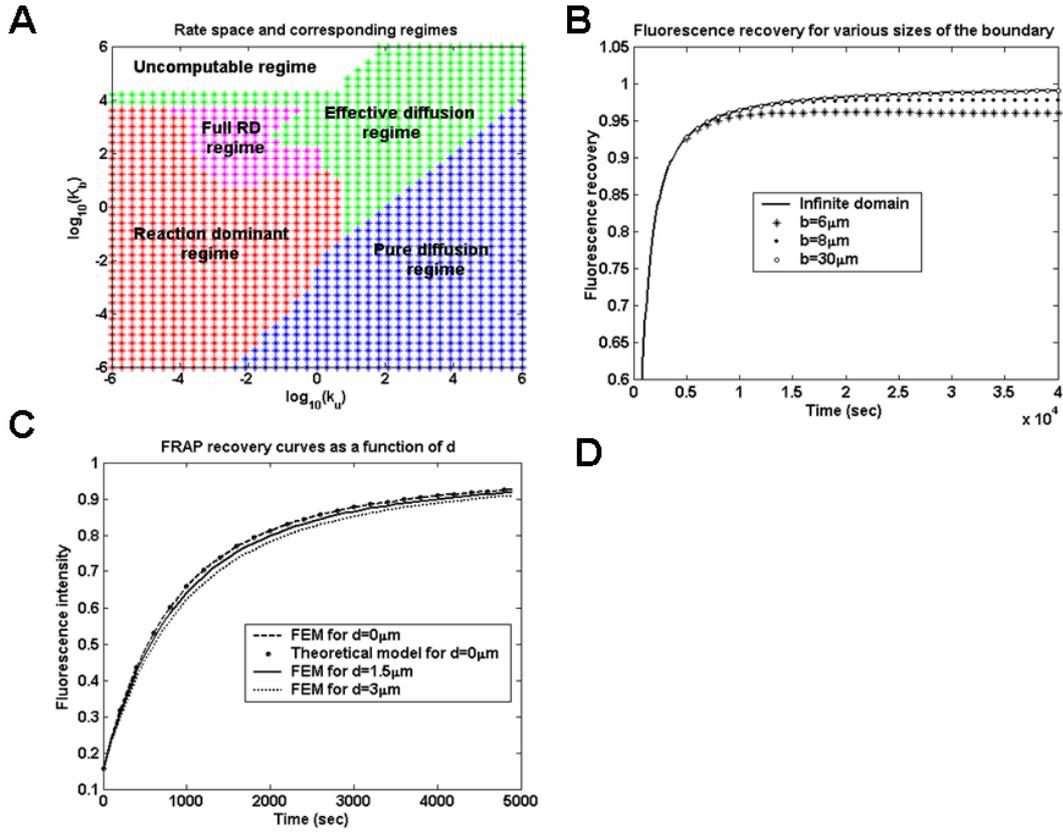

Figure 3: (A) Rate constant space and corresponding regimes for diffusion in a bounded domain. The sum of the squares of the residuals, *SQR*, was taken to be a criterion to draw the boundaries between the regions. The selected choice was *SQR*=0.1. (B) Fluorescence recovery curves for $K_b=10^2 \text{sec}^{-1}$, $k_u=10^{-2.5}\text{sec}^{-1}$, for three values of the size of the membrane boundary: (i) $b=6\mu m$ (*black* stars), (ii) $b=8\mu m$ (*black* dots) and (iii) $b=30\mu m$ (open circles dots). As the size increases, the system behaves as if it were not bounded (solid line). (C) Fluorescence recovery curves for three values of *d*: 0μm (dashed line), 1.5μm (solid line) and 3μm (dotted line) after using a Finite Element Method to solve reaction-diffusion equations. Dots in red are derived from our model from Eq.7. The values used for the rate constants were $K_b=10^2\text{sec}^{-1}$, $k_u=10^{-2.5}\text{sec}^{-1}$.

**Supplementary material**

*A   Solution of reaction-diffusion equations in infinite domains*

We start with the reaction-diffusion equations (Eq.4 in the main manuscript) and we introduce the variables $N_f$ and $N_c$ that represent the non-fluorescent concentrations of the population $F$ and $C$, respectively: $N_f = F_o - f$ and $N_c = C_o - c$, where $\gamma = K_b/k_u$ and $F_o \equiv 1/(1+\gamma)$ and $C_o \equiv \gamma/(1+\gamma)$ (see Eq.5 in the main manuscript) following a characteristic procedure for this kind of problems (Crank 1975; Sprague et al 2004). The set of equations in Eq.4 have the same form for the new variables with the exception of the initial conditions:

$$N_f(r, t=0) = F_o C_i \left[1 - \exp(-K \exp(-2r^2/r_0^2))\right]$$
$$N_c(r, t=0) = C_o C_i \left[1 - \exp(-K \exp(-2r^2/r_0^2))\right]$$
(SM.1)

The application of a Laplace transform $\overline{N}(r, p) = \int_0^\infty N(r,t) e^{-pt} dt$ on Eq.4 removes the time derivative of the reaction-diffusion equations resulting into

$$p \overline{N}_f = D_f \nabla^2 \overline{N}_f - K_b \overline{N}_f + k_u \overline{N}_c + N_f(r, t=0)$$
$$p \overline{N}_c = K_b \overline{N}_f - k_u \overline{N}_c + N_c(r, t=0)$$
(SM.2)

Then, we introduce the variables, $q$ and $V$ defined by

$$q^2 = \frac{p}{D_f}\left(1 + \frac{K_b}{p+k_u}\right)$$

$$V = \frac{F_o}{D_f}\left(1 + \frac{K_b}{p+k_u}\right)\left[1 - \exp(-K \exp(-2r^2/r_0^2))\right]$$
(SM.3)

and Eq.SM.2 simplifies to

$$D_f \nabla^2 \overline{N}_f - \overline{N}_f q^2 = -V$$
(SM.4)

We subsequently define a cut-off value $A$ for the radius at which $V$ has a very small value. Taking into account the specification of the problem, we choose this value close to four times the bleach spot size when essentially $V \approx 0$. There are two regions then, one in which $V \neq 0$ and a second where $V = 0$. To solution in every region can be calculated easily (Carslaw and Jaeger 1959), taking into account the continuity of the solution and its first derivative at $r = SM$. We are interested more in the form of solution inside the bleached spot (i.e. $r<A$). Eq.(SM.4) yields then the following expression for the concentration of the non-fluorescent molecules in the $p$-space for $r<A$:

$$\overline{N}_f(p) = -\frac{F_o}{D_f}\left(1+\frac{K_b}{p+k_u}\right)\sum_{n=1}^{\infty}\frac{(-K)^n}{n!}B(q,A,w,n,r_o),$$

$$B(q,A,w,n,r_o) = \left[\int drr K_o(qr)\exp(-2n(r/r_o)^2)\Big|_{r=A} I_o(qr) - \right. \quad (SM.5)$$

$$\left. I_o(qr)\int drr K_o(qr)\exp(-2n(r/r_o)^2) + K_o(qr)\int drr I_o(qr)\exp(-2n(r/r_o)^2)\right],$$

Fluorescence inside the bleached spot results from computation of the inverse Laplace transform of the average of the sum of fluorescent populations $F$ and $C$

$$Fluorescence(t) = InvLaplTransf\left(\left\langle \frac{1}{p}-\overline{N}_f-\overline{N}_c \right\rangle\right) \equiv InvLaplTransf(F(p)), \text{ where}$$

$$F(p) = \frac{1}{p} + \frac{2F_o}{w^2 D_f}\left(1+\frac{K_b}{p+k_u}\right)^2 \sum_{n=1}^{\infty}\frac{(-K)^n}{n!}B(q,A,w,n,r_o) +$$

$$\frac{C_o}{p+k_u}\frac{1}{2}\left(\frac{r_o}{w}\right)^2\sum_{n=1}^{\infty}\frac{(-K)^n}{n!n}\left(1-\exp(-2n(w/r_o)^2)\right),$$

$$B(q,A,w,n,r_o) = \left[\int_0^w drr\left(\int drr K_o(qr)\exp(-2n(r/r_o)^2)\right)\Big|_{r=A} I_o(qr) - \right.$$

$$\left.\int_0^w drr I_o(qr)\int drr K_o(qr)\exp(-2n(r/r_o)^2) + \int_0^w drr K_o(qr)\int drr I_o(qr)\exp(-2n(r/r_o)^2)\right],$$

$$q^2 = \frac{p}{D_f}\left(1+\frac{K_b}{p+k_u}\right)$$

(SM.6)

*B Solution of reaction-diffusion equations in bounded domains*

A similar procedure to that followed in the beginning of Section A is going to be ensued in the case for which there is a membrane that does not allow molecules to escape. The equations have to be modified to take into account the existence of a barrier at distance $r=b$ from the centre of the nucleus (Figure 1_SM).

Although, the same set of reaction-diffusion equations governs the dynamics of the system, Neumann boundary conditions should be introduced as constraints corresponding to the fact that the flux of fluorescent molecules outside the nuclear membrane

$$\left.\frac{\partial f}{\partial r}\right|_{r=b} = \left.\frac{\partial c}{\partial r}\right|_{r=b} = 0$$

while the initial conditions Eq.SM.1 hold the same as in the infinite domain case. The only introduction compared to the previous procedure is that concentration of fluorescent (and non-fluorescent) molecules outside the boundary should be zero. The concentration of the non-fluorescent molecules for $r<A$ (note that $A<b$) is given by the following expression

$$\overline{N}_f(p) = -\frac{F_0}{D_f}\left(1+\frac{Kb}{p+k_u}\right)\sum_{n=1}^{\infty}\frac{(-K)^n}{n!}B(q,A,w,n,r_0),$$

$$B(q,A,w,n,r_0) = \left[\int drr\left[K_0(qr)+I_0(qr)\frac{K_1(qb)}{I_1(qb)}\right]\exp(-2n(r/r_0)^2)\right]_{r=A} I_0(qr) - \text{(SM.7)}$$

$$I_0(qr)\int drrK_0(qr)\exp(-2n(r/r_0)^2) + K_0(qr)\int drrI_0(qr)\exp(-2n(r/r_0)^2)\right],$$

and the total fluorescence inside the bleached area (i.e. $r<w$) in the $p$-space is

$$F(p) = \frac{1}{p} + \frac{2F_0}{w^2 D_f}\left(1+\frac{Kb}{p+k_u}\right)^2\sum_{n=1}^{\infty}\frac{(-K)^n}{n!}B(q,A,w,n,r_0) +$$

$$\frac{C_0}{p+k_u}\frac{1}{2}\left(\frac{r_0}{w}\right)^2\sum_{n=1}^{\infty}\frac{(-K)^n}{n!n}\left(1-\exp(-2n(w/r_0)^2)\right),$$

$$B(q,A,w,n,r_0) = \left[\int_0^w drr\left(\int drr\left[K_0(qr)+I_0(qr)\frac{K_1(qb)}{I_1(qb)}\right]\exp(-2n(r/r_0)^2)\right)_{r=A} I_0(qr) - \right. \qquad \text{(SM.8)}$$

$$\int_0^w drrI_0(qr)\int drrK_0(qr)\exp(-2n(r/r_0)^2) + \int_0^w drrK_0(qr)\int drrI_0(qr)\exp(-2n(r/r_0)^2)\right],$$

$$q^2 = \frac{p}{D_f}\left(1+\frac{Kb}{p+k_u}\right)$$

Fluorescence dependence on time can then be calculated by using inverse Laplace transform.

*I. Pure diffusion reduction of reaction-diffusion equation*

For a two-dimensional system which exhibits a pure-diffusional behaviour, we introduce a small parameter $\varepsilon$ (<<1) with $K_b=\varepsilon K^* << k_u$. We consider an expansion of the variables $f$ and $c$ according to the expressions

$$f(r,t) = \sum_{n=0}^{\infty} \varepsilon^n f_n(r,t) \quad \text{and} \quad c(r,t) = \sum_{n=0}^{\infty} \varepsilon^n c_n(r,t) \quad (SM.9)$$

If we replace $f$ and $c$ with the expressions Eq.SM.9 and substitute them into Eq.4, equations reduce to (by keeping only the leading term)

$$\frac{\partial f_o(r,t)}{\partial t} = D_f \nabla_r^2 f_o(r,t) + k_u c_o(r,t),$$

$$\frac{\partial c_o(r,t)}{\partial t} = -k_u c_o(r,t)$$

for t>0 (SM.10)

with initial conditions

$$f_o(r,t=0) = \frac{1}{1+\gamma} C_i \exp(-K \exp(-2r^2/r_o^2))$$

$$c_o(r,t=0) = \frac{\gamma}{1+\gamma} C_i \exp(-K \exp(-2r^2/r_o^2))$$

Due to the fact that $\gamma<<1$, $c_o(r,t=0)\approx 0$. The solution of Eq.SM.10 gives for the total population, Total$\equiv f(r,t)+c(r,t)$

$$\frac{\partial Total(r,t)}{\partial t} = D_f \nabla_r^2 Total(r,t),$$

$$Total(r,t=0) \cong \exp(-K \exp(-2r^2/r_o^2))$$

(SM.11)

We note that for $\gamma<<1$, the system of the reaction-diffusion equations is reduced to a pure diffusion equation.

Pure diffusion expression in bounded domains can be derived by following a variable separation procedure (Crank 1975) for the spatial and temporal parts of the pure diffusion equation. For diffusion of fluorescent molecules in a cylindrical volume bounded at $r=b$, we obtain the following expression for the concentration of fluorescent material (Crank 1975)

$$f(r,t) = C_i - \frac{2}{b^2} \int_0^b F(r) r dr - \frac{2}{b^2} \sum_{n=1}^{\infty} \exp(-D_f t \alpha_n^2) \frac{J_o(\alpha_n r)}{(J_o(\alpha_n b))^2} \int_0^b r F(r) J_o(\alpha_n r) dr$$

(SM.12)

where $J_o$ are the zero-th order Bessel functions of the first kind and $\alpha_n$ are values for which the first derivative of the first order Bessel functions of the first kind, $J_1(\alpha_n b)$ equals zero (this results from the boundary condition). By recalling that $F(r)=C_{bleached}$, Eq.SM.12 turns into (noting that $J_1(\alpha_n b)=0$)

$$f(r,t) = \frac{r_o^2 C_i}{2b^2} \sum_{m=1}^{\infty} \frac{(-K)^m}{m m!}\left(1-\exp(-2mb^2/r_o^2)\right) +$$

$$\frac{2C_i}{b^2} \sum_{n=1}^{\infty} \exp(-D_f t \alpha_n^2) \frac{J_o(\alpha_n r)}{(J_o(\alpha_n b))^2} \int_0^b r J_o(\alpha_n r) \left[\sum_{m=1}^{\infty} \frac{(-K)^m}{m!} \exp(-2mr^2/r_o^2)\right] dr \qquad (SM.13)$$

and by integrating the concentration Eq.SM.13 over a circle of radius $w$ and dividing with the initial fluorescence in this disk we obtain the final formula for the FRAP recovery

$$F(t) = C_i + \frac{r_o^2 C_i}{2b^2} \sum_{m=1}^{\infty} \frac{(-K)^m}{m m!}\left(1-\exp(-2mb^2/r_o^2)\right) +$$

$$\frac{4C_i}{w^2 b^2} \int_0^w dr \sum_{n=1}^{\infty} \exp(-D_f t \alpha_n^2) \frac{J_o(\alpha_n r)}{(J_o(\alpha_n b))^2} \left\{\int_0^b r J_o(\alpha_n r) \left[\sum_{m=1}^{\infty} \frac{(-K)^m}{m!} \exp(-2mr^2/r_o^2)\right] dr\right\}$$

(SM.14)

Unlike the expression for pure diffusion in an infinite domain, formula Eq.SM.14 do not lead to a closed form for diffusion in a bounded domain and only a numerical computation is available.

*II. Effective diffusion reduction of reaction-diffusion equation*

Similarly, when both the wandering time between binding events and the residency time in the binding sites are very small and the reaction process is much faster than diffusion, we can introduce (Carrero et al 2004a) a small number $\varepsilon$ ($<<1$) and define $K_b = B/\varepsilon$ and $k_u = U/\varepsilon$. Then, if we add together Eqs.4, we get

$$\frac{\partial (f+c)(r,t)}{\partial t} = D_f \nabla_r^2 f(r,t),$$

$$\varepsilon \frac{\partial c(r,t)}{\partial t} = Bf(r,t) - Uc(r,t) \qquad \text{for } t>0 \qquad (SM.15)$$

After the substitution of Eq.SM.9 in Eq.SM.15 the leading-order term can be obtained and the equations for $f$ and $c$ uncouple

$$\frac{\partial f_o(r,t)}{\partial t} = \frac{D_f}{1+\gamma} \nabla_r^2 f_o(r,t) \quad \text{and} \quad \frac{\partial c_o(r,t)}{\partial t} = \frac{D_f}{1+\gamma} \nabla_r^2 c_o(r,t)$$

and for the whole population of fluorescent molecules

$$\frac{\partial Total(r,t)}{\partial t} = D_{eff} \nabla_r^2 Total(r,t) \qquad (SM.16)$$

where $D_{eff} = \frac{Df}{1+\gamma}$. As a result, for a pair of rate constants in the effective diffusion regime, the concentration of fluorescent molecules is similar to formula describing molecules that diffuse freely with the replacement of the diffusion coefficient with an effective one, reduced by the term $1/(1+\gamma)$.

*III. Reaction-dominant reduction of reaction-diffusion equation*

When the system lies in the reaction-dominant regime, diffusion of molecules is a very fast process. As a result, it is the concentration of fluorescent bound molecules that has a time-dependence. By contrast, the concentration of free molecules reaches fast an equilibrium. The total number of free fluorescent molecules inside the circle of radius $b$ is $\pi b^2 f$, where $f$ is the concentration of free fluorescent molecules inside the nucleus. This number equals to the number of molecules that are not bleached

$$\int_0^b dr\, 2\pi r F_o C_i \exp(-K \exp(-2r^2/r_o^2)) = \pi b^2 f \Rightarrow$$

$$f = F_o C_i \left[ 1 + \sum_{n=1}^{\infty} \frac{(-K)^n}{2n!\,n} \left(\frac{r_o}{b}\right)^2 \left(1 - \exp(-2nb^2/r_o^2)\right) \right] \qquad (SM.17)$$

If we substitute this expression into the second equation of Eq.4, we obtain the concentration of the fluorescent bound molecules

$$c(t) = C_o C_i + \frac{C_o C_i}{2}\left(\frac{r_o}{b}\right)^2 \sum_{n=1}^{\infty} \frac{(-K)^n}{n!\,n}\left(1 - \exp(-2n(b/r_o)^2)\right) +$$

$$C_o C_i \left[ \frac{1}{2}\left(\frac{r_o}{w}\right)^2 \sum_{n=1}^{\infty} \frac{(-K)^n}{n!\,n}\left(1 - \exp(-2n(w/r_o)^2)\right) - \right. \qquad (SM.18)$$

$$\left. \frac{1}{2}\left(\frac{r_o}{b}\right)^2 \sum_{n=1}^{\infty} \frac{(-K)^n}{n!\,n}\left(1 - \exp(-2n(b/r_o)^2)\right) \right] \exp(-k_u t)$$

and if we recall that $F_o+C_o=1$ and $C_o=\gamma/(1+\gamma)$, we finally get

$$F(t) = 1 + \frac{1}{2}\left(\frac{r_o}{b}\right)^2 \sum_{n=1}^{\infty} \frac{(-K)^n}{n!n}\left(1 - \exp(-2n(b/r_o)^2)\right) +$$

$$\frac{\gamma}{1+\gamma}\left[\frac{1}{2}\left(\frac{r_o}{w}\right)^2 \sum_{n=1}^{\infty} \frac{(-K)^n}{n!n}\left(1 - \exp(-2n(w/r_o)^2)\right) - \right.$$

$$\left. \frac{1}{2}\left(\frac{r_o}{b}\right)^2 \sum_{n=1}^{\infty} \frac{(-K)^n}{n!n}\left(1 - \exp(-2n(b/r_o)^2)\right)\right] \exp(-k_u t) \qquad \text{(SM.19)}$$

*C Finite element method for solving reaction-diffusion equations for off-centre bleached spots*

To estimate the fluorescence recovery when there is a displacement of the bleached spot, the system of reaction-diffusion equations Eq.4 (in the main manuscript) will still be valid but with the replacement of the Laplacian operator $\nabla_r^2 \rightarrow \nabla_{x,y}^2$. Without loss of generality and for the sake of a simpler formulation, we have assumed that the bleached spot has been displaced along only the *x*-direction. We have chosen a bleached spot is displaced at *d*=4μm from the nucleus centre (Figure 2_SM A). To the best of our knowledge, an analytical expression cannot be obtained and a numerical calculation of the reaction-diffusion equations was performed with a finite element method that is employed by the use of the powerful and flexible *pdetool* of Matlab: A refined mesh with 2145 nodes and 4160 triangles inside the nucleus was used (Figure 2_SM B).

**List of Figures**

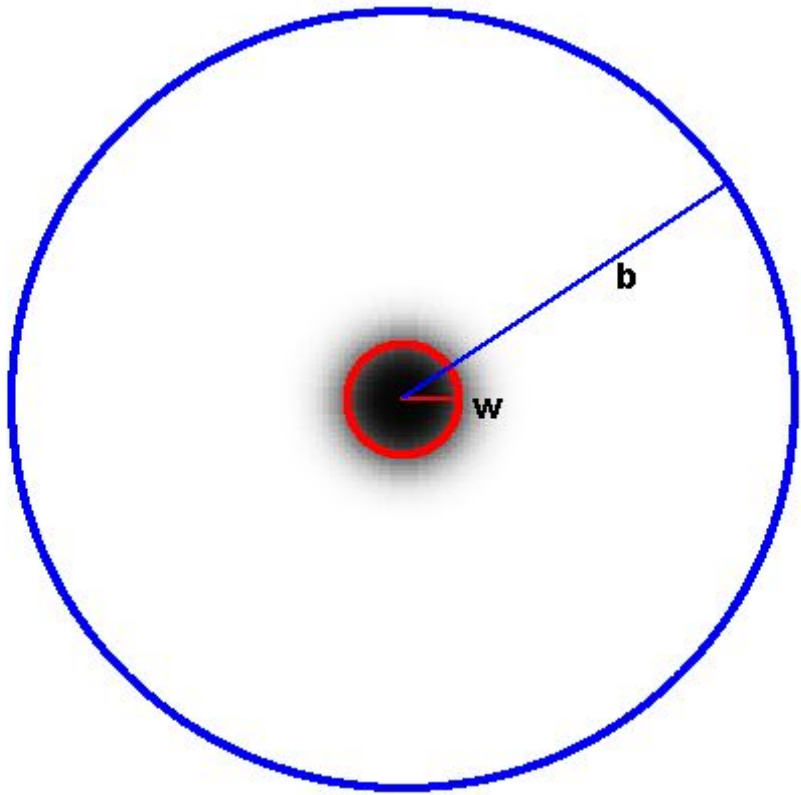

Figure 1_SM: Fluorescence profile after bleaching a circular spot of radius *w* for diffusion inside a bounded domain of radius *b*.

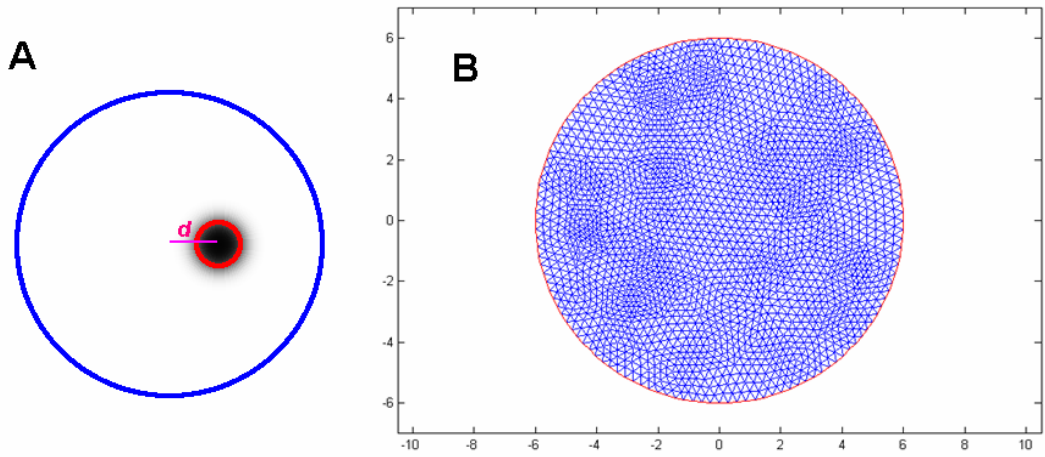

Figure 2_SM: (A) Fluorescence profile after bleaching a circular spot at a distance *d* from the centre of the nucleus. (B) Mesh used to solve the reaction diffusion equations using a Finite Element Method.